\documentclass[letterpaper, 10pt]{IEEEtran}

\usepackage{cite}
\usepackage{amsmath,amssymb,amsfonts}
\usepackage{algorithmic}
\usepackage{graphicx}
\usepackage{textcomp}
\usepackage{xcolor}
\usepackage{mathrsfs}
\usepackage{psfrag}
\usepackage{soul}

\definecolor{mcl}{HTML}{000000}

\def\BibTeX{{\rm B\kern-.05em{\sc i\kern-.025em b}\kern-.08em
    T\kern-.1667em\lower.7ex\hbox{E}\kern-.125emX}}
\begin{document}
\title{\LARGE \bf Output Feedback Consensus for Networked Heterogeneous Nonlinear Negative-Imaginary Systems with Free Body Motion\\
}

\author{Kanghong Shi, \quad Ian R. Petersen, \IEEEmembership{Fellow, IEEE}, \quad and \quad Igor G. Vladimirov
\thanks{This work was supported by the Australian Research Council under grant DP190102158.}
\thanks{K. Shi, I. R. Petersen and I. G. Vladimirov are with the School of Engineering, College of Engineering and Computer Science, Australian National University, Canberra, ACT 2600, Australia. (email: kanghong.shi@anu.edu.au; ian.petersen@anu.edu.au; igor.vladimirov@anu.edu.au)%
}
}

\newtheorem{definition}{Definition}
\newtheorem{theorem}{Theorem}
\newtheorem{conjecture}{Conjecture}
\newtheorem{lemma}{Lemma}
\newtheorem{remark}{Remark}
\newtheorem{corollary}{Corollary}

\maketitle

\thispagestyle{empty}
\pagestyle{empty}

\begin{abstract}
This paper provides a protocol to address the robust output feedback consensus problem for networked heterogeneous nonlinear negative-imaginary (NI) systems with free body dynamics. {\color{mcl}We extend the definition of nonlinear NI systems to allow for systems with free body motion. A new stability result is developed for the interconnection of a nonlinear NI system and a nonlinear output strictly negative-imaginary (OSNI) system. Also, a class of networked nonlinear OSNI controllers is proposed to achieve output feedback consensus for heterogeneous networked nonlinear NI systems. We show that in this control framework, the system outputs converge to the same limit trajectory. This consensus protocol is illustrated by a numerical example.}

\end{abstract}


\section{INTRODUCTION}
Negative-Imaginary (NI) systems theory was introduced by Lanzon and Petersen in \cite{lanzon2008stability} and \cite{petersen2010feedback} to address a robust control problem for flexible structures and has attracted much attention among control theory researchers (see \cite{bhikkaji2011negative,xiong2009lossless,song2012negative,xiong2012finite,song2012robust,wang2015robust,xiong2010negative}). Typical NI systems are systems with colocated force actuators and position sensors. Positive-Real (PR) systems theory \cite{brogliato2007dissipative} cannot be applied to the control of such systems in general. An NI system was initially defined in \cite{lanzon2008stability} to be a stable system with a frequency response $F(j\omega)$ satisfying $j(F(j\omega)-F(j\omega)^*)\geq 0$ for all $\omega>0$. Examples of such systems arise in lightly damped structures \cite{bhikkaji2009fast,fanson1990positive,cai2010stability} and nano-positioning systems \cite{van2010modal,devasia2007survey,sebastian2005design,dong2007robust}. The stability of an NI system with transfer function matrix $F(s)$ can be guaranteed by applying a strictly Negative-Imaginary (SNI) controller with transfer function matrix $G(s)$ such that the DC gain condition $\lambda_{max}(F(0)G(0))<1$ is satisfied \cite{lanzon2008stability}.

The definition of NI systems in \cite{lanzon2008stability,petersen2010feedback} was extended in  \cite{xiong2010negative} to include systems with poles in the closed left half of the complex plane except at the origin. The definition has been extended again in \cite{mabrok2014generalizing} to include systems with poles at the origin. Systems with free body motion such as single integrators and double integrators were included in this new definition and a new robust stability result for NI systems in \cite{mabrok2014generalizing}. The original definition of NI systems has also been recently extended to include nonlinear systems \cite{ghallab2018extending} and some stability results were established for nonlinear NI systems in \cite{ghallab2018extending} and \cite{shi2020robust}. However, systems with free body motion are excluded in the nonlinear NI definition in \cite{ghallab2018extending}.

{\color{mcl}
Cooperative control for multi-agent systems has been a highly active research area over the past two decades \cite{murray2007recent}. This control paradigm enables multiple systems to perform team missions and has been applied on autonomous vehicles including mobile robots, unmanned air vehicles (UAVs), autonomous underwater vehicles (AUVs) and other applications \cite{ren2005survey}. Consensus, that is convergence to an agreement through information sharing among agents, is one of the most important problems in the area of cooperative control \cite{ren2007consensus}. Consensus algorithms were first studied for first-order dynamics (see \cite{jadbabaie2003coordination,olfati2004consensus,moreau2005stability}, etc) and then extended to second-order systems (see \cite{tanner2003stable,saber2003flocking,yu2013distributed,su2011adaptive}, etc).

As NI properties arise naturally in autonomous vehicles and a wide variety of other applications \cite{petersen2016negative}, consensus problems were investigated in \cite{wang2015robust} for heterogeneous NI systems. Using NI systems theory, the consensus algorithms in \cite{wang2015robust} only require outputs of the agents, and consensus is guaranteed if a simple DC gain condition is satisfied. The theoretical result presented in \cite{wang2015robust} has already been applied to real-world cooperative control problems (see \cite{tran2017formation,qi2016cooperative,hu2019distributed,peng2016robust,skeik2019cooperative}). However, this result is restricted to linear systems. Motivated by nonlinear NI systems theory, the result of \cite{wang2015robust} has been recently extended in \cite{shi2020robustb}, which provides a protocol that achieves output feedback consensus for networked heterogeneous nonlinear NI systems. However, \cite{shi2020robustb} uses the definition of nonlinear NI systems from \cite{ghallab2018extending}, which excludes systems with free body motion such as integrators.

In this work, we use an alternative definition for a class of nonlinear NI systems to include systems with free body motion. We obtain a stability result for the interconnection of a nonlinear NI system and a nonlinear output strictly negative-imaginary (OSNI) system (see also \cite{bhowmick2017lti} and \cite{bhowmick2019output} for linear OSNI systems), where nonlinear OSNI systems are also redefined to allow direct feedthrough from the input to the output. Then we consider the output feedback consensus problem for multiple nonlinear NI systems with different state-space models. We aim to find a control protocol such that the difference between the outputs of any two agents connected in a network converges to zero. We model the communication between multiple nonlinear NI systems using an undirected connected graph, where the nonlinear NI plants and the nonlinear OSNI controllers correspond to the nodes and the edges of the graph, respectively. Each controller takes the difference between the outputs of the two plants connected to it as input and feeds back its output to the plants. This forms an augmented system whose stability is investigated in order to achieve output consensus for the nonlinear NI plants. We prove the networked nonlinear NI plants act as an augmented nonlinear NI system and the networked nonlinear OSNI controllers act as an augmented nonlinear OSNI system. Therefore, the entire networked control system can be regarded as the interconnection of a nonlinear NI system and a nonlinear OSNI system. Similarly to the stability result mentioned above, an extended stability result is established for the networked control system that guarantees output consensus. This shows that for networked heterogeneous nonlinear NI systems satisfying certain assumptions, output feedback consensus can be achieved by using suitable nonlinear OSNI controllers in the proposed control framework. The control framework is robust with respect to bounded perturbations in the system models for both the nonlinear NI plants and the nonlinear OSNI controllers.

This paper contributes an output feedback consensus protocol for networked heterogeneous nonlinear NI systems with free body motion. This work differs from the related previous results in the following aspects: \cite{ghallab2018extending} defines nonlinear NI systems but does not allow for systems with free body motion while this work does; \cite{shi2020robust} only considers consensus for agents with the same model while this work allows agents to have different state-space models; \cite{shi2020robust} and \cite{shi2020robustb} only deal with consensus problems for nonlinear NI systems without free body motion while this restriction is lifted in this work. This work can also be regarded as an extension of the papers \cite{mabrok2014generalizing} and \cite{wang2015robust} to nonlinear systems.}

Notation: The notation in this paper is standard. $\mathbb R$ and $\mathbb C$ denote the fields of real and complex numbers, respectively. $\mathbb R^{m\times n}$ and $\mathbb C^{m\times n}$ denote the spaces of real and complex matrices of dimension $m\times n$, respectively. $A^T$ and $A^*$ denote the transpose and complex conjugate transpose of a matrix $A$, respectively. $\lambda_{max}(\cdot)$ denotes the maximum eigenvalue of a matrix with only real eigenvalues. {\color{mcl}$A\otimes B$ denotes the Kronecker product of matrices $A$ and $B$.} {\color{mcl}$\bar{\cdot}$ denotes a constant value for a given vector or scalar signal.} {\color{mcl}$\|\cdot \|$ denotes the Euclidean norm of a vector.} $I_n$ is the $n\times n$ identity matrix. For a nonlinear dynamical system $H$ with input $u$ and output $y$, $y=H(u)$ describes its input-output relationship. $diag\{a_1,a_2,\cdots,a_l\}$ represents a diagonal matrix with the values $a_1,a_2,\cdots,a_l$ on its diagonal.

Graph Theory Preliminaries: $\mathcal G=(\mathcal V,\mathcal E)$, where $\mathcal V=\{v_1,v_2,\cdots,v_N\}$ and $\mathcal E=\{e_1,e_2,\cdots,e_l\} \subseteq \mathcal V\times \mathcal V$, describes an undirected graph with $N$ nodes and $l$ edges. The corresponding symmetric adjacency matrix $\mathcal A = [a_{ij}]\in \mathbb R^{N\times N}$ is defined so that $a_{ii}=0$, and $\forall i\neq j$, $a_{ij}=1$ if $(v_i,v_j)\in \mathcal E$ and $a_{ij}=0$ otherwise. A sequence of unrepeated edges in $\mathcal E$ that joins a sequence of nodes in $\mathcal V$ defines a path. An undirected graph is connected if there is a path between every pair of nodes. Given an undirected graph $\mathcal G$, a corresponding directed graph can be obtained by defining a direction for each edge of $\mathcal G$. The incidence matrix $\mathcal Q=[q_{ev}]\in \mathbb{R}^{l\times N}$ of a directed graph is defined so that the elements in $\mathcal Q$ are given by
\begin{equation*}
    q_{ev}:=
    \begin{cases}
    1 & \text{if } v \text{ is the initial vertex of edge } e,\\
    -1 & \text{if } v \text{ is the terminal vertex of edge } e,\\
    0 & \text{if } v \text{ does not belong to edge } e.
    \end{cases}
\end{equation*}
In this paper, the initial and terminal vertices of an edge in a directed graph can both send information to each other via the corresponding controller. For an undirected graph $\mathcal G$, the choice of a corresponding directed graph is not unique. However, the Laplacian matrix $\mathcal L_N$ of $\mathcal G$ has the following relationship with the incidence matrix $\mathcal Q$ of any directed graph corresponding to $\mathcal G$: $\mathcal{L}_N=\mathcal{Q}^T\mathcal{Q}$. 
\section{AN INITIAL STABILITY RESULT}
\label{sec:initial stability}
In this section, new definitions of nonlinear NI and nonlinear OSNI systems are provided and a new stability result is established for the interconnection of a single nonlinear NI system and a single nonlinear OSNI system. The new stability result is applicable to nonlinear NI systems with free body motion, which are excluded in the previous stability result given in \cite{shi2020robust}.

Consider the following general nonlinear system:
\begin{align}
    \dot x(t)=&\ f(x(t),u(t)),\label{eq:state equation of nonlinear NI}\\
    y(t)=&\ h(x(t))+Du(t),\label{eq:output equation of nonlinear NI}
\end{align}
where {\color{mcl}$x(t)\in \mathbb R^{n}$ is the state, $u(t)\in \mathbb R^m$ is the input, and $y(t)\in \mathbb R^m$ is the output,} $f:\mathbb R^n\times \mathbb R^m \to \mathbb R^n$ is a Lipschitz continuous function, $h:\mathbb R^n \to \mathbb R^m$ is a class $C^1$ function and $D\in \mathbb R^{m\times m}$ is a symmetric matrix; i.e., $D=D^T$.
\begin{definition}\label{def:nonlinear NI}
The system (\ref{eq:state equation of nonlinear NI}), (\ref{eq:output equation of nonlinear NI}) is said to be a nonlinear negative-imaginary (NI) system if there exists a positive semidefinite storage function $V:\mathbb R^n\to \mathbb R$ of class $C^1$ such that
\begin{equation}\label{eq:NI MIMO definition inequality}
    \dot V(x(t))\leq u(t)^T\dot {\tilde y}(t),
\end{equation}
for all $t\geq 0$, where
\begin{equation}\label{eq:tilde y}
	\tilde y(t)=h(x(t)).
\end{equation}
\end{definition}

In contrast to Definition 3 in \cite{ghallab2018extending}, which excludes linear NI systems with poles at the origin, Definition \ref{def:nonlinear NI} now includes all linear NI systems satisfying the definition given in \cite{mabrok2014generalizing} by allowing the storage function of the system to be positive semidefinite instead of positive definite.
\begin{definition}\label{def:nonlinear OSNI}
The system (\ref{eq:state equation of nonlinear NI}), (\ref{eq:output equation of nonlinear NI}) is said to be a nonlinear output strictly negative-imaginary (OSNI) system if there exists a positive semidefinite storage function $V:\mathbb R^n\to\mathbb R$ of class $C^1$ and a {\color{mcl}scalar} $\epsilon>0$ such that
\begin{equation}\label{eq:dissipativity of OSNI}
    \dot V(x(t))\leq u(t)^T\dot {\tilde y}(t) -\epsilon \left\|\dot {\tilde y}(t)\right\|^2,
\end{equation}
for all $t\geq 0$, where $\tilde y(t)$ is as defined in (\ref{eq:tilde y}). In this case, we also say that system (\ref{eq:state equation of nonlinear NI}), (\ref{eq:output equation of nonlinear NI}) is nonlinear OSNI with degree of output strictness $\epsilon$.
\end{definition}

In this paper, nonlinear OSNI systems defined as in Definition \ref{def:nonlinear OSNI} are applied as controllers to achieve the robust output feedback consensus for networked heterogeneous nonlinear NI systems defined as in Definition \ref{def:nonlinear NI}. First, we provide a stability result for a single feedback interconnection of nonlinear NI systems. 

Consider a multiple-input multiple-output (MIMO) nonlinear NI system $H_1$ with the following state-space model: 
\begin{align}
   H_1: \quad \dot x_1(t)=&\ f_1(x_1(t),u_1(t)),\label{eq:state equation of H1}\\
    y_1(t)=&\ h_1(x_1(t)),\label{eq:output equation of H1}
\end{align}
where {\color{mcl}$x_1(t)\in \mathbb R^{n}$ is the state, $u_1(t)\in \mathbb R^m$ is the input, and $y_1(t)\in \mathbb R^m$ is the output,} $f_1:\mathbb R^n\times \mathbb R^m \to \mathbb R^n$ is a Lipschitz continuous function and $h_1:\mathbb R^n \to \mathbb R^m$ is a class $C^1$ function.

For the system $H_1$ with the state-space model (\ref{eq:state equation of H1}), (\ref{eq:output equation of H1}), we suppose the following assumption is satisfied.

{\bf Assumption I}: When the system $H_1$ is in steady state; i.e.,  $u_1(t)\equiv \bar u_1$, $x_1(t)\equiv \bar x_1$ and $y_1(t)\equiv \bar y_1$, we have $\bar u_1^T \bar y_1\geq 0$.

For nonlinear NI systems, Assumption I corresponds to the property of linear NI systems stated in Lemma 2 in \cite{lanzon2008stability}.

Consider a MIMO nonlinear OSNI system $H_2$ with the following state-space model: 
\begin{align}
   H_2: \quad \dot x_2(t)=&\ f_2(x_2(t),u_2(t)),\label{eq:state equation of H2}\\
    y_2(t)=&\ h_2(x_2(t))+D_2u_2(t),\label{eq:output equation of H2}
\end{align}
where {\color{mcl}$x_2(t)\in \mathbb R^{n}$ is the state, $u_2(t)\in \mathbb R^m$ is the input, and $y_2(t)\in \mathbb R^m$ is the output,} $f_2:\mathbb R^n\times \mathbb R^m \to \mathbb R^n$ is a Lipschitz continuous function, $h_2:\mathbb R^n \to \mathbb R^m$ is a class $C^1$ function and $D_2\in \mathbb R^{m\times m}$ is a symmetric matrix; i.e., $D_2=D_2^T$.

For the system $H_2$ with the state-space model (\ref{eq:state equation of H2}), (\ref{eq:output equation of H2}), we suppose that the following assumption is satisfied.

{\bf Assumption II}: When the system $H_2$ is in steady state; i.e.,  $u_2(t)\equiv \bar u_2$, $x_2(t)\equiv \bar x_2$ and $y_2(t)\equiv \bar y_2$, we have $\bar u_2^T \bar y_2 \leq -\gamma \|\bar u_2\|^2$ with $\gamma > 0$.

It might be observed that as nonlinear OSNI systems belong to a subclass of nonlinear NI systems, Assumption II seems to have a conflicting relationship with Assumption I. In fact, Assumption II can be satisfied because of the  term $D_2u_2(t)$ in the output equation (\ref{eq:output equation of H2}) and it corresponds to the inequality (61) for linear NI systems in \cite{mabrok2014generalizing}.

In addition, both of the systems $H_1$ and $H_2$ are assumed to satisfy the following assumptions. For the system $H_1$ with input $u_1(t)$, state $x_1(t)$ and output $y_1(t)=h_1(x_1(t))$ described by the state-space model (\ref{eq:state equation of H1}), (\ref{eq:output equation of H1}) and the system $H_2$ with input $u_2(t)$, state $x_2(t)$ and the auxiliary output $\tilde y_2(t)=h_2(x_2(t))$ described by the state-space model (\ref{eq:state equation of H2}), (\ref{eq:output equation of H2}), we suppose for $i=1$ and $2$, the following conditions are satisfied.

{\bf Assumption III}: Over any time interval $[t_a,t_b]$ where $t_b>t_a$, $h_i(x_i(t))$ remains constant if and only if $x_i(t)$ remains constant; i.e., $\dot h_i(x_i(t))\equiv 0\iff \dot x_i(t)\equiv 0$. Moreover, $h_i(x_i(t))\equiv 0 \iff x_i(t)\equiv 0$.

{\bf Assumption IV}: Over any time interval $[t_a,t_b]$ where $t_b>t_a$, $x_i(t)$ remains constant only if $u_i(t)$ remains constant; i.e., $x_i(t)\equiv \bar x_i \implies u_i(t)\equiv \bar u_i $. Moreover, $x_i(t)\equiv 0 \implies u_i(t)\equiv 0$.

In the case of linear systems, Assumption III corresponds to observability and Assumption IV corresponds to the $B$ matrix in the realisation $(A,B,C,D)$ of the linear system having full column rank.

\begin{figure}[h!]
\centering
\psfrag{H_1}{$H_1$}
\psfrag{in_0}{$w=0$}
\psfrag{y_1}{\hspace{0.07cm}$y_1$}
\psfrag{in_2}{$u_{2}$}
\psfrag{y_2}{$y_{2}$}
\psfrag{in_1}{$u_{1}$}
\psfrag{H_2}{$H_2$}
\hspace{0.5cm}\includegraphics[width=8cm]{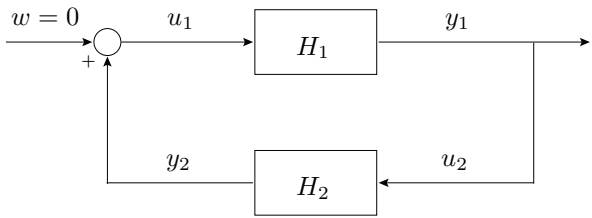}
\caption{Closed-loop interconnection of a MIMO nonlinear NI system $H_1$ and a MIMO nonlinear OSNI system $H_2$.}
\label{fig:single closed-loop}
\end{figure}

\begin{theorem}\label{theorem:single}
Consider the closed-loop positive feedback interconnection of the system $H_1$ with state-space model (\ref{eq:state equation of H1}), (\ref{eq:output equation of H1}) and $H_2$ with state-space model (\ref{eq:state equation of H2}), (\ref{eq:output equation of H2}), as shown in Fig.~\ref{fig:single closed-loop}. Suppose that Assumptions I-IV are satisfied, and the storage function, defined as
\begin{align}
    W(x_1,x_2):=& \ V_1(x_1)+V_2(x_2)-h_1(x_1)^Th_2(x_2)\notag\\
    &-\frac{1}{2}h_1(x_1)^TD_2h_1(x_1)\label{eq:W},
\end{align}
is positive definite, where $V_1(x_1)$ and $V_2(x_2)$ are positive semidefinite storage functions that satisfy (\ref{eq:NI MIMO definition inequality}) for the system $H_1$ and (\ref{eq:dissipativity of OSNI}) for the system $H_2$, respectively. Then, the closed-loop interconnection of the systems $H_1$ and $H_2$ is asymptotically stable.
\end{theorem}
\begin{IEEEproof}
According to the nonlinear NI property (\ref{eq:NI MIMO definition inequality}) for the system $H_1$, the nonlinear OSNI property (\ref{eq:dissipativity of OSNI}) for the system $H_2$ and the system setting $u_1(t)\equiv y_2(t)$ and $u_2(t)\equiv y_1(t)$ in Fig.~\ref{fig:single closed-loop}, we have
\begin{align}
\dot V_1(x_1)\leq & \ u_1^T\dot y_1\notag\\
=&\ y_2^T\dot y_1\notag\\
=& [h_2(x_2)+D_2u_2]^T\dot h_1(x_1)\notag\\
=& [h_2(x_2)+D_2y_1]^T\dot h_1(x_1)\notag\\
=& [h_2(x_2)+D_2h_1(x_1)]^T\dot h_1(x_1),\label{eq:NI property single theorem}
\end{align}
and
\begin{align}
\dot V_2(x_2)\leq & \ u_2^T\dot {\tilde y}_2-\epsilon \|\dot {\tilde y}_2\|^2\notag\\
=&\ y_1^T\dot {\tilde y}_2-\epsilon \|\dot {\tilde y}_2\|^2\notag\\
=&\ h_1(x_1)^T\dot h_2(x_2)-\epsilon \|\dot h_2(x_2)\|^2,\label{eq:OSNI property single theorem}
\end{align}
{\color{mcl}where the above equalities also use (\ref{eq:output equation of H1}) and (\ref{eq:output equation of H2}).} We obtain the time derivative of the storage function $W(x_1,x_2)$ in (\ref{eq:W}) using (\ref{eq:NI property single theorem}) and (\ref{eq:OSNI property single theorem}):
\begin{align}
	\dot W(x_1,x_2)=&\ \dot V_1(x_1)+\dot V_2(x_2)-\dot h_1(x_1)^Th_2(x_2)\notag\\
	&-h_1(x_1)^T\dot h_2(x_2)
	-h_1(x_1)^TD_2\dot h_1(x_1)\notag\\
	\leq &\left[h_2(x_2)+D_2h_1(x_1) \right]^T\dot h_1(x_1)\notag\\
	&+h_1(x_1)^T\dot h_2(x_2)-\epsilon \|\dot h_2(x_2)\|^2\notag\\
	&-\dot h_1(x_1)^Th_2(x_2)-h_1(x_1)^T\dot h_2(x_2)\notag\\
	&-h_1(x_1)^TD_2\dot h_1(x_1)\notag\\
	=&-\epsilon \|\dot h_2(x_2)\|^2\notag\\
	\leq &\ 0.\label{eq:W_dot in single theorem}
\end{align}
From this it follows that $\dot W(x_1,x_2)=0$ is only possible when $\dot h_2(x_2)=0$. Hence, $\dot W(x_1,x_2)$ can remain zero only if $\dot h_2(x_2)$ remains zero; i.e., $\dot W(x_1,x_2)\equiv 0\implies \dot h_2(x_2(t))\equiv 0$. According to Assumptions III and IV, $\dot h_2(x_2(t))\equiv 0\implies \dot x_2(t)\equiv 0\implies u_2(t)\equiv \bar u_2$. Hence, the system $H_2$ is in steady-state. According to the system setting in Fig.~\ref{fig:single closed-loop}, $y_1(t)\equiv u_2(t)$. Hence, using Assumptions III and IV, $\dot y_1(t)\equiv 0\implies\dot x_1(t)\equiv 0\implies u_1(t)\equiv \bar u_1$. Thus, the system $H_1$ is also in steady-state. Then, according to Assumption II, we have
\begin{equation*}
\bar u_2^T\bar y_2\leq -\gamma \|\bar u_2\|^2.
\end{equation*}
If $\bar u_2=0$, then $\bar u_2^T\bar y_2=0$. According to Assumptions III and IV, and the system setting in Fig.~\ref{fig:single closed-loop}, $\bar u_2=0\implies \bar y_1=0\implies \bar x_1=0 \implies \bar u_1=0 \implies \bar y_2=0 \implies \bar x_2 = 0$. Hence, in this case, the system is in equilibrium. Otherwise, if $\bar u_2\neq 0$, we have
\begin{equation}\label{eq:u_2y_2 inequality}
	\bar u_2^T\bar y_2<0.
\end{equation}
Also, according to Assumption I, we have
\begin{equation}\label{eq:u_1y_1 inequality}
	\bar u_1^T\bar y_1\geq 0.
\end{equation}
According to the system setting in Fig.~\ref{fig:single closed-loop}, we have $\bar u_1=\bar y_2$ and $\bar y_1=\bar u_2$. Hence, (\ref{eq:u_1y_1 inequality}) can be rewritten as
\begin{equation*}
	\bar u_2^T\bar y_2 \geq 0,
\end{equation*}
which contradicts (\ref{eq:u_2y_2 inequality}). Thus, we can conclude that $\dot W(x_1,x_2)$ cannot remain zero unless $x_1=x_2=0$. Thus, according to LaSalle's invariance principle, $W(x_1,x_2)$ will keep decreasing until $W(x_1,x_2)=0$. Hence, the equilibrium at $(x_1,x_2)=(0,0)$ of the closed-loop interconnection is asymptotically stable.
\end{IEEEproof}

\section{OUTPUT FEEDBACK CONSENSUS}
\label{sec:consensus}
\begin{figure}[h!]
\centering
\psfrag{H_0}{$\mathcal{H}_p$}
\psfrag{u_p1}{$u_{p1}$}
\psfrag{y_p1}{$y_{p1}$}
\psfrag{u_p2}{$u_{p2}$}
\psfrag{y_p2}{$y_{p2}$}
\psfrag{u_pn}{$u_{pN}$}
\psfrag{y_pn}{$y_{pN}$}
\psfrag{H_p1}{\hspace{-0.05cm}$H_{p1}$}
\psfrag{H_p2}{\hspace{-0.05cm}$H_{p2}$}
\psfrag{H_pn}{\hspace{-0.05cm}$H_{pN}$}
\psfrag{ddd}{\hspace{0.13cm}$\vdots$}
\psfrag{udd}{\hspace{0.15cm}$\vdots$}
\psfrag{odd}{\hspace{0.1cm}$\vdots$}
\hspace{0.5cm}\includegraphics[width=8cm]{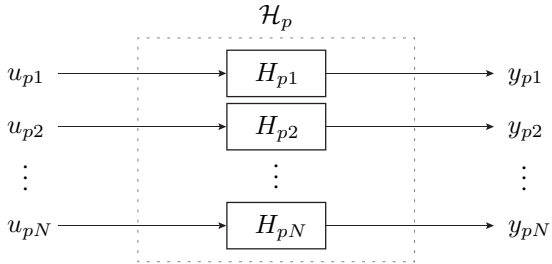}
\caption{A plant $\mathcal{H}_p$ consisting of $N$ independent and heterogeneous nonlinear NI systems $H_{pi}$ $(i=1,2,\cdots,N)$ in (\ref{eq:state pi}) and (\ref{eq:output pi}), with independent inputs and outputs combined as the input and output of the networked system $\mathcal{H}_p$.}
\label{fig:H_p1_hetero}
\end{figure}

Consider $N$ heterogeneous nonlinear plants $H_{pi}$ $(i=1,2,\cdots,N)$ described as
\begin{align}
  H_{pi}: \quad  \dot x_{pi}(t)=&\ f_{pi}(x_{pi}(t),u_{pi}(t)),\label{eq:state pi}\\
    y_{pi}(t)=&\ h_{pi}(x_{pi}(t)),\label{eq:output pi}
\end{align}
where {\color{mcl}$x_{pi}(t)\in \mathbb R^{n}$ is the state, $u_{pi}(t)\in \mathbb R^m$ is the input, and $y_{pi}(t)\in \mathbb R^m$ is the output,} $f_{pi}:\mathbb R^n \times \mathbb R^m \to \mathbb R^n$ are Lipschitz continuous functions and $h_{pi}:\mathbb R^n \to \mathbb R^m$ are class $C^1$ functions. These systems operate independently in parallel and each of them has its own input $u_{pi}\in \mathbb{R}^m$ and output $y_{pi}\in\mathbb{R}^m$, ($i=1,2,\cdots,N$), which is shown in Fig.~\ref{fig:H_p1_hetero}. The subscript ``$p$" indicates that this system will play the role of a plant in what follows. We combine the inputs and outputs respectively as the vectors $U_p=[u_{p1}^T,u_{p2}^T,\cdots,u_{pN}^T]^T\in \mathbb R^{Nm\times 1}$ and $Y_p=[y_{p1}^T,y_{p2}^T,\cdots,y_{pN}^T]^T=[h_{p1}(x_{p1})^T,h_{p2}(x_{p2})^T,\cdots,h_{pN}(x_{pN})^T]^T\in \mathbb R^{Nm\times 1}$, respectively.

\begin{figure}[h!]
\centering
\psfrag{H_p}{$\mathcal{H}_p$}
\psfrag{H_p1}{$H_{p1}$}
\psfrag{H_p2}{$H_{p2}$}
\psfrag{H_pl}{$H_{pN}$}
\psfrag{U_p}{\hspace{-0.05cm}$U_{p}$}
\psfrag{Y_p}{\hspace{0.08cm}$Y_{p}$}
\psfrag{ddots}{$\ddots$}
\psfrag{H_np}{$\hat{\mathcal{H}}_{p}$}
\psfrag{U_np}{\hspace{0.2cm}$\hat{U}_{p}$}
\psfrag{Y_np}{\hspace{0.2cm}$\hat{Y}_{p}$}
\psfrag{Q_t}{\hspace{0.05cm}$\mathcal{Q}\otimes I_m$}
\psfrag{Q}{\hspace{-0.17cm}$\mathcal{Q}^T\otimes I_m$}
\includegraphics[width=9cm]{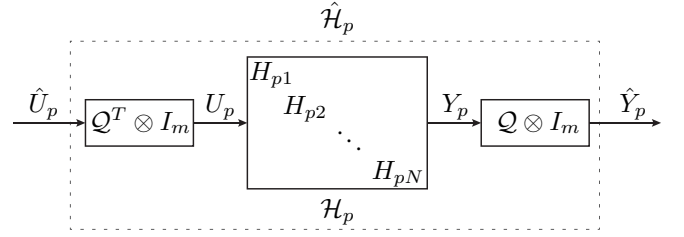}
\caption{Heterogeneous nonlinear NI plants connected according to the directed graph network topology.}
\label{fig:networked_plants}
\end{figure}

Let us consider the networked plants connected according to the graph network topology $\hat{\mathcal{H}}_{p}$ as shown in Fig.~\ref{fig:networked_plants}, where $\mathcal{Q}$ is the incidence matrix of a directed graph that represents the communication links between the heterogeneous nonlinear NI plants.

For the system $\hat{\mathcal H}_p$ shown in Fig.~\ref{fig:networked_plants}, we have the following lemma:
\begin{lemma}
If the plants $H_{pi}$ are nonlinear NI systems for all $i=1,2,\cdots,N$, then the networked plant $\hat{\mathcal H}_p$ is also a nonlinear NI system.
\end{lemma}
\begin{IEEEproof}
According to Definition \ref{def:nonlinear NI}, each nonlinear NI system $H_{pi}$ ($i=1,2,\cdots,N$) must have a corresponding positive semidefinite storage function $V_{pi}(x_{pi})$ such that $\dot V_{pi}(x_{pi})\leq u_{pi}^T\dot y_{pi}$, where $x_{pi}$ is the state of the system $H_{pi}$. We define the storage function for the system $\hat{\mathcal{H}}_p$ as $\hat V_p=\sum_{i=1}^N V_{pi}(x_{pi})$, which is positive semidefinite. Then
\begin{equation}\label{eq:V_p ineq}
    \dot{\hat V}_p=\sum_{i=1}^N \dot V_{pi}(x_{pi})\leq \sum_{i=1}^N u_{pi}^T\dot y_{pi}=U_p^T\dot Y_p.
\end{equation}
Let $\hat U_p$ and $\hat Y_p$ denote the input and output of the system $\hat{\mathcal{H}}_{p}$, respectively. According to the system setting in Fig.~\ref{fig:networked_plants}, we have
\begin{equation*}
U_p=(\mathcal Q^T\otimes I_m) \hat U_p,\quad \textnormal{and} \quad \hat Y_p=(\mathcal Q\otimes I_m) Y_p.
\end{equation*}
Therefore, we have
\begin{equation}\label{eq:H_p and hat H_p i-o relation}
U_p^TY_p=[(\mathcal Q^T\otimes I_m) \hat U_p)]^TY_p=\hat U_p^T(\mathcal Q\otimes I_m) Y_p=\hat U_p^T\hat Y_p.
\end{equation}
According to (\ref{eq:V_p ineq}) and (\ref{eq:H_p and hat H_p i-o relation}), we obtain the nonlinear NI inequality for the system $\hat {\mathcal H}_p$:
\begin{equation}\label{eq:hat V_p dot}
\dot{\hat V}_p \leq \hat U_p^T\dot{\hat Y}_p.	
\end{equation}
Therefore, $\hat{\mathcal{H}}_p$ is a nonlinear NI system.
\end{IEEEproof}

Now we give a definition of output feedback consensus for a network of systems as shown in Fig.~\ref{fig:H_p1_hetero}.
\begin{definition}
A distributed output feedback control law achieves output feedback consensus for a network of systems if $|y_{pi}(t) - y_{pj}(t)|\to 0$ as $t\to +\infty$, $\forall i,j\in\{1,2,\cdots,N\}$.
\end{definition}

\begin{figure}[h!]
\centering
\psfrag{H_0}{$\mathcal{H}_c$}
\psfrag{u_p1}{$u_{c1}$}
\psfrag{y_p1}{$y_{c1}$}
\psfrag{u_p2}{$u_{c2}$}
\psfrag{y_p2}{$y_{c2}$}
\psfrag{u_pn}{$u_{cl}$}
\psfrag{y_pn}{$y_{cl}$}
\psfrag{H_p1}{\hspace{-0.05cm}$H_{c1}$}
\psfrag{H_p2}{\hspace{-0.05cm}$H_{c2}$}
\psfrag{H_pn}{\hspace{-0.05cm}$H_{cl}$}
\psfrag{ddd}{\hspace{0.13cm}$\vdots$}
\psfrag{udd}{\hspace{0.15cm}$\vdots$}
\psfrag{odd}{\hspace{0.1cm}$\vdots$}
\hspace{0.5cm}\includegraphics[width=8cm]{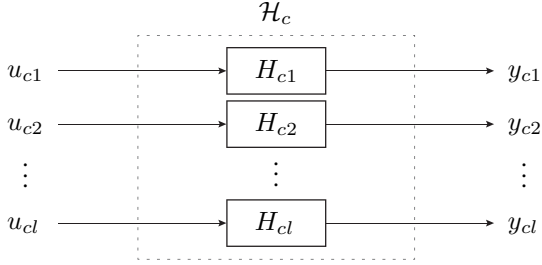}
\caption{A controller $\mathcal{H}_c$ consisting of $l$ independent and heterogeneous nonlinear OSNI systems $H_{ck}$ $(k=1,2,\cdots,l)$ in (\ref{eq:state ck}) and (\ref{eq: output ck}), with independent inputs and outputs combined as the input and output of the networked system $\mathcal{H}_c$.}
\label{fig:H_c1_hetero}
\end{figure}

Consider a series of heterogeneous nonlinear OSNI controllers $H_{ck}$ $(k=1,2,\cdots,l)$ applied at the edges in the network. The OSNI controllers have the following state-space models:
\begin{align}
  H_{ck}: \quad  \dot x_{ck}(t)=&\ f_{ck}(x_{ck}(t),u_{ck}(t)),\label{eq:state ck}\\
    y_{ck}(t)=&\ h_{ck}(x_{ck}(t))+D_{ck}u_{ck}(t)\label{eq: output ck},
\end{align}
where {\color{mcl}$x_{ck}(t)\in \mathbb R^{q}$ is the state, $u_{ck}(t)\in \mathbb R^m$ is the input, and $y_{ck}(t)\in \mathbb R^m$ is the output,} $f_{ck}:\mathbb R^q \times \mathbb R^m \to \mathbb R^q$ are Lipschitz continuous functions, $h_{ck}:\mathbb R^q \to \mathbb R^m$ are class $C^1$ functions and $D_{ck}\in \mathbb R^{m\times m}$ are symmetric matrices. These systems operate independently in parallel and each of them has its own input $u_{ck}\in \mathbb{R}$ and output $y_{ck}\in\mathbb{R}$, $k=1,2,\cdots,l$, which is shown in Fig.~\ref{fig:H_c1_hetero}. The subscript ``$c$" indicates that this system will play the role of a controller in what follows. We combine the inputs and outputs respectively as the vectors $U_c=[u_{c1}^T,u_{c2}^T,\cdots,u_{cl}^T]^T\in \mathbb R^{lm\times 1}$ and $Y_c=[y_{c1}^T,y_{c2}^T,\cdots,y_{cl}^T]^T=\Pi_c+D_cU_c\in \mathbb R^{lm\times 1}$, where
\begin{equation}\label{eq:Pi def}
\Pi_c=\left[\begin{matrix}h_{c1}(x_{c1})\\h_{c2}(x_{c2})\\\vdots \\ h_{cl}(x_{cl})\end{matrix}\right]\in \mathbb R^{lm\times1},
\end{equation}
and
\begin{equation}\label{eq:J def}
	D_c=diag\{D_{c1},D_{c2},\cdots,D_{cl}\}\in \mathbb R^{lm\times lm}.
\end{equation}

\begin{lemma}
If the controllers $H_{ck}$ are nonlinear OSNI systems for all $k=1,2,\cdots,l$, then the networked controller $\mathcal{H}_{c}$ is also a nonlinear OSNI system.
\end{lemma}
\begin{IEEEproof}
For every nonlinear OSNI system $H_{ck}$, we have a positive semidefinite storage function $V_{ck}(x_{ck})$ and a constant $\epsilon_k>0$ such that
\begin{equation}\label{eq:OSNI ineq for controllers}
    \dot V_{ck}(x_{ck})\leq u_{ck}^T\dot {\tilde y}_{ck}-\epsilon_k \|\dot {\tilde y}_{ck}\|^2,
\end{equation}
where $\tilde y_{ck}=h_{ck}(x_{ck})$ and $\epsilon_k$ is the level of output strictness of the system $H_{ck}$. For the system $\mathcal{H}_c$, we define its storage function $V_c$ as the sum of the storage functions of all the networked controllers; i.e.,$V_c := \sum_{k=1}^l V_{ck}(x_{ck})$, which is positive semidefinite. The time derivative of $V_c$ is:
\begin{align}
    \dot V_c =& \sum_{k=1}^l \dot V_{ck}(x_{ck})\notag\\
    \leq & \sum_{k=1}^l u_{ck}^T\dot {\tilde y}_{ck}-\sum_{k=1}^l\epsilon_k \|\dot {\tilde y}_{ck}\|^2\notag\\
    \leq & \sum_{k=1}^l u_{ck}^T\dot {\tilde y}_{ck}-\epsilon_{min}\sum_{k=1}^l\|\dot {\tilde y}_{ck}\|^2\notag\\
  =&\ U_c^T \dot \Pi_c-\epsilon_{min}\|\dot \Pi_c\|^2,\label{eq:hat V_c ineq}
\end{align}
where $\epsilon_{min}=min\{\epsilon_1,\epsilon_2,\cdots,\epsilon_l\}$. Hence, the system $\mathcal{H}_c$ satisfies the definition of a nonlinear OSNI system and $\epsilon_{min}$ quantifies a level of output strictness of the system. This completes the proof.
\end{IEEEproof}

\begin{figure}[h!]
\centering
\psfrag{H_p}{$\mathcal{H}_p$}
\psfrag{H_c}{$\mathcal{H}_c$}
\psfrag{H_p1}{$H_{p1}$}
\psfrag{H_p2}{$H_{p2}$}
\psfrag{H_pn}{$H_{pN}$}
\psfrag{H_c1}{$H_{c1}$}
\psfrag{H_c2}{$H_{c2}$}
\psfrag{H_cl}{$H_{cl}$}
\psfrag{U_p}{$U_{p}$}
\psfrag{Y_p}{$Y_{p}$}
\psfrag{U_c}{$U_{c}$}
\psfrag{Y_c}{$Y_{c}$}
\psfrag{ddots}{$\ddots$}
\psfrag{H_np}{$\hat {\mathcal H}_p$}
\psfrag{U_np}{\hspace{0.1cm}$\hat U_p$}
\psfrag{Y_np}{\hspace{0.1cm}$\hat Y_p$}
\psfrag{Q}{\hspace{-0.22cm}$\mathcal{Q}^T\otimes I_m$}
\psfrag{Q_t}{\hspace{0.05cm}$\mathcal{Q}\otimes I_m$}
\includegraphics[width=9cm]{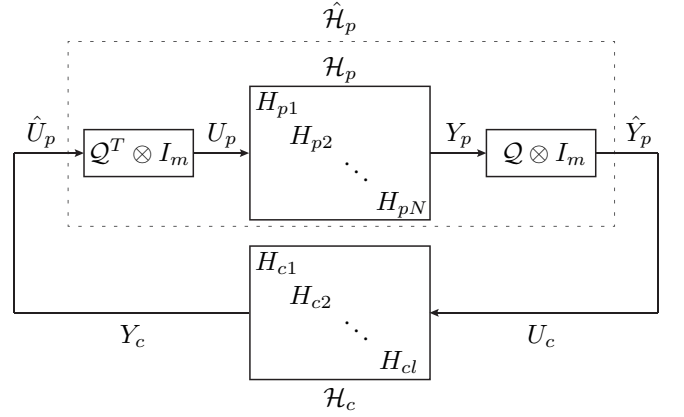}
\caption{Positive feedback interconnection of heterogeneous nonlinear NI plants and nonlinear OSNI controllers according to the directed graph network topology.}
\label{fig:closed_network}
\end{figure}

Now consider the closed-loop positive feedback interconnection of the networked plants shown in Fig.~\ref{fig:networked_plants} and the networked controllers shown in Fig.~\ref{fig:H_c1_hetero}, which is depicted in Fig.~\ref{fig:closed_network}. In this paper, robust output consensus of heterogeneous nonlinear NI plants is achieved by constructing a control system with the block diagram shown in Fig.~\ref{fig:closed_network} and choosing suitable controllers that satisfy certain conditions. Detailed description of the control framework in Fig.~\ref{fig:closed_network} can be found in \cite{shi2020robustb}, which uses a similar control framework.

{\color{mcl}As is shown in Fig.~\ref{fig:closed_network}, the control input for the networked plant $\mathcal H_p$ is
\begin{equation}\label{eq:protocol}
    U_p = (\mathcal{Q}^T\otimes I_m)\mathcal{H}_c\left((\mathcal{Q}\otimes I_m)Y_p\right),
\end{equation}
where $\mathcal Q$ is the incidence matrix of the directed graph that represents the communication links between the heterogeneous nonlinear NI plants. Equivalently, the distributed control protocol for the plant $H_i \ (i=1,2,\cdots,N)$ is defined by the equations
\begin{align}  
    \dot x_{ck}(t)=&\ f_{ck}\left(x_{ck}(t),\sum_{j=1}^N q_{kj}y_{pj}\right),\label{eq:distributed x}\\
    y_{ck}(t)=&\ h_{ck}(x_{ck}(t))+D_{ck}\sum_{j=1}^N q_{kj}y_{pj}\label{eq:distributed y},\\
    u_{pi}=&\sum_{k=1}^l q_{ki}y_{ck},\label{eq:distributed u}
\end{align}
where $q_{kj}$ and $q_{ki}$ are the $j$-th and $i$-th elements in the $k$-th row of the incidence matrix $\mathcal Q$, respectively. Here, $\sum_{j=1}^N q_{kj}y_{pj}$ represents the difference between the outputs of the two plants connected by the edge $e_k$.}

Before we present the main result, let us make another assumption. For the system $\hat{\mathcal{H}}_p$ with input $\hat U_p(t)$ and output $\hat Y_p(t)$, we suppose the following assumption is satisfied.

{\bf Assumption V}: Given a constant input $\hat U_p(t)\equiv \bar{\hat U}_p$ to the system $\hat{\mathcal{H}}_p$, if its output is also constant; i.e., $\hat Y_p(t)\equiv\bar{\hat Y}_p$, then $\bar{\hat U}_p$ and $\bar{\hat Y}_p$ satisfy $\bar{\hat U}_p^T\bar{\hat Y}_p\geq 0$.

In fact, Assumption I implies Assumption V in the case that all the plants $H_{pi}$ are in steady state. This is because when all the plants satisfy Assumption I and are in steady state, we have $\bar U_p^T\bar Y_p\geq 0$ and according to the system setting in Fig.~\ref{fig:closed_network} we have $\bar U_p^T\bar Y_p=[(\mathcal Q^T\otimes I_m)^T\bar{\hat U}_p]^T\bar Y_p=\bar{\hat U}_p^T(\mathcal Q\otimes I_m)\bar Y_p=\bar{\hat U}_p^T\bar{\hat Y}_p$ similarly to (\ref{eq:H_p and hat H_p i-o relation}). Hence $\bar{\hat U}_p^T\bar{\hat Y}_p\geq 0$. However, Assumption V is assumed for the networked plants $\hat{\mathcal H}_p$ in the following theorem instead of assuming Assumption I for each individual plant because Assumption V also allows for the situation in which the input and output of the system $\hat{\mathcal H}_p$ are constant, but the individual plants $H_{pi}$ are not all in steady state. This situation is possible because the matrix $\mathcal Q\otimes I_m$ takes the difference between the outputs of the plants. Under constant inputs, if the plants oscillate with a constant difference between their outputs, then this situation is allowed under Assumption V.

\begin{theorem}\label{theorem:consensus}
Consider an undirected connected graph $\mathcal G$ that models the communication links for a network of heterogeneous nonlinear NI systems $H_{pi}$ $(i=1,2,\cdots,N)$ as shown in Fig.~\ref{fig:H_p1_hetero}, and any directed graph corresponding to $\mathcal G$ with the incidence matrix $\mathcal Q$. Also, consider the heterogeneous nonlinear OSNI control laws $H_{ck}$ $(k=1,2,\cdots,l)$ for all of the edges. Suppose Assumptions III and IV are satisfied for the plants $H_{pi}$, Assumptions II, III and IV are satisfied for the controllers $H_{ck}$ (with $\gamma=\gamma_k$ for the controller $H_{ck}$ in Assumption II) and Assumption V is satisfied by the system $\hat{\mathcal{H}}_p$. Also, suppose the storage function, defined as
\begin{equation*}
    \hat W:=\hat V_p+V_c-\hat Y_p^T\Pi_c -\frac{1}{2}\hat Y_p^T D_c \hat Y_p ,
\end{equation*}
is positive definite, where $\hat V_p$ and $V_c$ are positive semidefinite storage functions that satisfy (\ref{eq:hat V_p dot}) for the system $\hat{\mathcal{H}}_p$ and (\ref{eq:hat V_c ineq}) for the system $\hat {\mathcal{H}}_c$, respectively. Here, $\hat Y_p$ is the output of the system $\hat{\mathcal{H}}_p$. $\Pi_c$ and $D_c$ are terms in the output $Y_c$ of the system $\mathcal{H}$ and are defined in (\ref{eq:Pi def}) and (\ref{eq:J def}). Then robust output feedback consensus can be achieved via the protocol (\ref{eq:protocol}),
or equivalently (\ref{eq:distributed x})-(\ref{eq:distributed u}) in a distributed manner for each plant $p_i$, as shown in Fig.~\ref{fig:closed_network}.
\end{theorem}
\begin{IEEEproof}
According to (\ref{eq:hat V_p dot}), (\ref{eq:hat V_c ineq}) and the system setting $\hat U_p \equiv Y_c$ and $ U_c \equiv \hat Y_p$ shown in Fig.~\ref{fig:closed_network}, we have
\begin{equation}\label{eq:dot V_p in proof}
    \dot {\hat V}_p \leq \hat U_p^T\dot{\hat Y}_p = Y_c^T\dot {\hat Y}_p=\dot {\hat Y}_p^T[\Pi_c+D_c U_c]=\dot {\hat Y}_p^T[\Pi_c+ D_c \hat Y_p],
\end{equation}
and
\begin{equation}\label{eq:dot hat V_c in proof}
\dot {V}_c\leq U_c^T\dot{\Pi}_c-\epsilon_{min} \|\dot \Pi_c\|^2=\hat Y_p^T\dot{\Pi}_c-\epsilon_{min} \|\dot \Pi_c\|^2.
\end{equation}
According to (\ref{eq:dot V_p in proof}), (\ref{eq:dot hat V_c in proof}) and the symmetry of $D_c$ in (\ref{eq:J def}), the time derivative of the storage function $\hat W$ satisfies the following inequality:
\begin{align}
    \dot {\hat W}=&\ \dot {\hat V}_p+\dot {V}_c-\dot {\hat Y}_p^T\Pi_c - \hat Y_p^T\dot{\Pi}_c-\frac{1}{2}\dot {\hat Y}_p^T(D_c+D_c^T)\hat Y_p\notag\\
    \leq &\ \dot {\hat Y}_p^T[\Pi_c+ D_c \hat Y_p] + \hat Y_p^T\dot{\Pi}_c-\epsilon_{min} \|\dot \Pi_c\|^2-\dot {\hat Y}_p^T\Pi_c \notag\\
    &- \hat Y_p^T\dot{\Pi}_c -\dot {\hat Y}_p^TD_c \hat Y_p\notag\\
    \leq &\ \epsilon_{min} \|\dot \Pi_c\|^2\notag\\
    \leq &\ 0.\label{eq:W dot}
\end{align}
Hence, the closed-loop system is at least Lyapunov stable. Moreover, $\dot {\hat W}=0$ can hold only if $\dot \Pi_c=0$. In other words, $\dot {\hat W}$ can remain zero only if $\dot h_{ck}(x_{ck}(t))$ remains zero for all $k=1,2,\cdots,l$. According to Assumptions III and IV, $\dot h_{ck}(x_{ck}(t))\equiv 0\implies\dot x_{ck}(t)\equiv 0\implies u_{ck}(t)\equiv \bar u_{ck}$. Hence, $H_{ck}$ is in steady-state for all $k=1,2,\cdots,l$. We have $U_c(t)\equiv \bar U_c$ and $Y_c(t)\equiv \bar Y_c$. According to the system setting in Fig.~\ref{fig:closed_network} that $\hat U_p(t)\equiv Y_c(t)$ and $U_c(t)\equiv \hat Y_p(t)$, we also have $\hat U_p(t)\equiv \bar{\hat U}_p$ and $\hat Y_p(t)\equiv \bar {\hat Y}_p$. According to Assumption V, we have
\begin{equation}\label{eq:H_p condition in proof}
\bar {\hat U}_p^T\bar {\hat Y}_p \geq 0.
\end{equation}
According to Assumption II, we have
\begin{equation*}
\bar U_c^T\bar Y_c= \sum_{k=1}^l \bar u_{ck}^T\bar y_{ck}\leq -\sum_{k=1}^l \gamma_k \|\bar u_{ck}\|^2 \leq -\gamma_{min} \|\bar U_c\|^2,
\end{equation*}
where $\gamma_{min}=\min\{\gamma_1,\gamma_2\cdots,\gamma_l\}$. In the case that $\bar U_c \neq 0$, we have
\begin{equation}\label{eq:H_c condition in proof}
	\bar U_c^T\bar Y_c<0
\end{equation}
which contradicts (\ref{eq:H_p condition in proof}) because $\bar U_c^T\bar Y_c=\bar {\hat U}_p^T\bar {\hat Y}_p$. In the case that $\bar U_c=0$, all connected plants have the difference between their system outputs being zero. Hence, output consensus has already been achieved. Otherwise, $\dot {\hat W}$ cannot remain zero. According to LaSalle's invariance principle, $\hat W$ will keep decreasing until either $\bar U_c=0$ or $\hat W=0$. Thus, output consensus is achieved in both cases. This completes the proof.
\end{IEEEproof}

\section{ILLUSTRATIVE EXAMPLE}
{\color{mcl}
\begin{figure}[h!]
\centering
\psfrag{1}{\hspace{-0.02cm}\Large$v_1$}
\psfrag{2}{\hspace{0.0cm}\Large$v_2$}
\psfrag{3}{\hspace{-0.005cm}\Large$v_3$}
\psfrag{4}{\hspace{-0.005cm}\Large$v_4$}
\psfrag{e_1}{\Large$e_1$}
\psfrag{e_2}{\Large$e_2$}
\psfrag{e_3}{\Large$e_3$}
\psfrag{e_4}{\Large$e_4$}
\includegraphics[width=5cm]{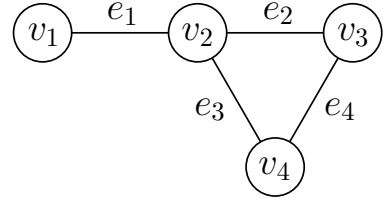}
\caption{An undirected and connected graph consisting of four nodes.}
\label{fig:3_nodes}
\end{figure}
Consider four nonlinear NI plants $H_{pi}$ at the vertices $v_i$ of the graph in Fig.~\ref{fig:3_nodes}. We choose directions of the edges as $e_1=(v_1,v_2)$, $e_2=(v_2,v_3)$, $e_3=(v_2,v_4)$ and $e_4=(v_3,v_4)$. Then the incidence matrix of the directed graph corresponding to $\mathcal G$ is
\begin{equation*}
	\mathcal Q = \left[\begin{matrix}1&-1&0&0\\0&1&-1&0\\0&1&0&-1\\0&0&1&-1 \end{matrix}\right].
\end{equation*}
The plants are nonlinear single integrators which have the following state-space models:
\begin{align*}
H_{pi}:\quad\dot x_{pi}(t)=&\ \mu_i u_{pi}^3(t),\\
y_{pi}(t)=&\ x_{pi}(t),\quad i=1,2,\cdots,4,
\end{align*}
where $\mu_1=10$, $\mu_2=30$, $\mu_3=25$ and $\mu_4=5$ are constant coefficients. The storage functions for these four plants are all $V_{pi}(x_{pi})=0$. The states $x_{p1}$, $x_{p2}$, $x_{p3}$ and $x_{p4}$ of these four plants have initial values $30$, $15$, $-5$ and $-10$, respectively. We aim to synchronise the outputs of these four plants to the same limit trajectory by using nonlinear OSNI controllers $H_{ck}$ at the edges $e_k$ of the graph in Fig.\ref{fig:3_nodes}.
\begin{align*}
H_{ck}:\quad \dot x_{ck}(t)=& -\alpha_k x_{ck}(t)-\beta_k x_{ck}^3(t)+u_{ck}(t),\\
y_{ck}(t)=&\ x_{ck}(t)-u_{ck}(t),\quad k = 1,2,\cdots,4,	
\end{align*}
where $\alpha_k$ and $\beta_k$ are constant coefficients. $\alpha_1 = 5$, $\alpha_2 = 8$, $\alpha_3 = 7$, $\alpha_4 = 3$; $\beta_1=3$, $\beta_2=2$, $\beta_3=5$ and $\beta_4=2$, respectively.

The storage functions of Controllers $1$, $2$, $3$ and $4$ are $V_{c1}=\frac{5}{2}x_{c1}^2+\frac{3}{4}x_{c1}^4$, $V_{c2}=4x_{c2}^2+\frac{1}{2}x_{c2}^4$, $V_{c3}=\frac{7}{2}x_{c3}^2+\frac{5}{4}x_{c3}^4$ and $V_{c4}=\frac{3}{2}x_{c4}^2+\frac{1}{2}x_{c4}^4$, respectively.

The networked plant system $\hat{\mathcal H}_p$ in this example is given by the equations:
\begin{align*}
\dot{\hat x}_1=&\ 10 \hat u_1^3+30(\hat u_1-\hat u_2-\hat u_3)^3,\\
\dot{\hat x}_2=&\ 30(-\hat u_1+\hat u_2+\hat u_3)^3+25(\hat u_2-\hat u_4)^3,\\
\dot{\hat x}_3=&\ 30(-\hat u_1+\hat u_2+\hat u_3)^3+5(\hat u_3+\hat u_4)^3,\\
\dot{\hat x}_4=&\ 25(-\hat u_2+\hat u_4)^3+5(\hat u_3+\hat u_4)^3,\\
\hat y_1=&\ \hat x_1,\quad 
\hat y_2= \hat x_2,\quad 
\hat y_3= \hat x_3,\quad
\hat y_4= \hat x_4.
\end{align*}

The storage function of the closed-loop system provided by Theorem \ref{theorem:consensus} takes the form
\begin{align*}
\hat W=&\	\frac{5}{2}x_{c1}^2+\frac{3}{4}x_{c1}^4+4x_{c2}^2+\frac{1}{2}x_{c2}^4+\frac{7}{2}x_{c3}^2+\frac{5}{4}x_{c3}^4+\frac{3}{2}x_{c4}^2\\
& +\frac{1}{2}x_{c4}^4 
-\hat x_1x_{c1}-\hat x_2x_{c2}-\hat x_3x_{c3}-\hat x_4x_{c4}+\frac{1}{2}\hat x_1^2\\
&+\frac{1}{2}\hat x_2^2+\frac{1}{2}\hat x_3^2+\frac{1}{2}\hat x_4^2\\
=& \left[\begin{matrix}\hat x_1&x_{c1}\end{matrix}\right]\left[\begin{matrix}\frac{1}{2}&-\frac{1}{2}\\-\frac{1}{2}&\frac{5}{2}\end{matrix}\right]\left[\begin{matrix}\hat x_1\\x_{c1}\end{matrix}\right]+\frac{3}{4}x_{c1}^4\\
&+\left[\begin{matrix}\hat x_2 & x_{c2}\end{matrix}\right]\left[\begin{matrix}\frac{1}{2}&-\frac{1}{2}\\-\frac{1}{2}&4\end{matrix}\right]\left[\begin{matrix}\hat x_2\\x_{c2}\end{matrix}\right]+\frac{1}{2}x_{c2}^4\\
&+\left[\begin{matrix}\hat x_3 & x_{c3}\end{matrix}\right]\left[\begin{matrix}\frac{1}{2}&-\frac{1}{2}\\-\frac{1}{2}&\frac{7}{2}\end{matrix}\right]\left[\begin{matrix}\hat x_3\\x_{c3}\end{matrix}\right]+\frac{5}{4}x_{c3}^4\\
&+\left[\begin{matrix}\hat x_4 & x_{c4}\end{matrix}\right]\left[\begin{matrix}\frac{1}{2}&-\frac{1}{2}\\-\frac{1}{2}&\frac{3}{2}\end{matrix}\right]\left[\begin{matrix}\hat x_4\\x_{c4}\end{matrix}\right]+\frac{1}{2}x_{c4}^4,
\end{align*}
which is positive definite. Assumptions II-V in Sections \ref{sec:initial stability} and \ref{sec:consensus} are satisfied. Output feedback consensus is achieved, as shown in Fig.~\ref{fig:example_figure3}. Because of the cubic nonlinearity in the plants and the controllers, their outputs have different rates of convergence in the domains where the states are far and close to the limit values. Therefore, a log scale is used for the time axis in the plot shown in Fig.~\ref{fig:example_figure3}.

\begin{figure}[h!]
\centering
\psfrag{Output}{\hspace{-0.1cm}Output}
\psfrag{Time (s)}{\hspace{-0.2cm}Time (s)}
\psfrag{Output Feedback Consensus}{\hspace{-0.3cm}Output Feedback Consensus}
\psfrag{0}{\scriptsize$0$}
\psfrag{10}{\scriptsize$10$}
\psfrag{20}{\scriptsize$20$}
\psfrag{30}{\scriptsize$30$}
\psfrag{-10}{\hspace{-0.13cm}\scriptsize$-10$}
\psfrag{e-6}{\scriptsize$10^{-6}$}
\psfrag{e-4}{\scriptsize$10^{-4}$}
\psfrag{e-2}{\scriptsize$10^{-2}$}
\psfrag{e0}{\scriptsize$10^{0}$}
\psfrag{e2}{\scriptsize$10^{2}$}

\psfrag{Plant 1 sp}{\scriptsize{Plant 1}}
\psfrag{Plant 2}{\scriptsize{Plant 2}}
\psfrag{Plant 3}{\scriptsize{Plant 3}}
\psfrag{Plant 4}{\scriptsize{Plant 4}}
\includegraphics[width=9cm]{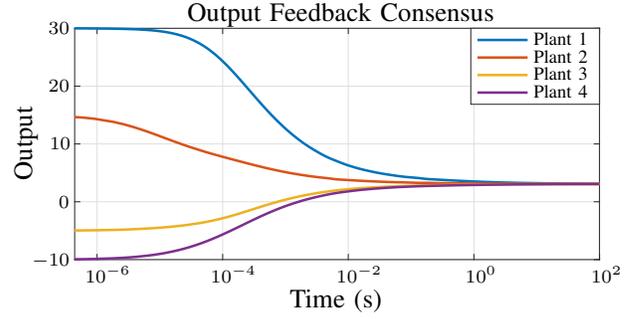}
\caption{Output feedback consensus for four nonlinear single integrator plants with different system models. {\color{mcl}Starting from different initial conditions, the outputs of the plants converge to the same limit trajectory under the effect of the proposed control framework of Section \ref{sec:consensus}.}}
\label{fig:example_figure3}
\end{figure}
}

\section{CONCLUSION}
This paper provides a control framework to achieve robust output feedback consensus for networked heterogeneous nonlinear NI systems including systems with free body motion, using nonlinear OSNI controllers. New definitions for nonlinear NI systems and nonlinear OSNI systems are given and a stability result is established for the simple feedback interconnection of a nonlinear NI plant and a nonlinear OSNI controller. A networked control framework is then considered by modelling the communication topology between systems as a connected graph, where the plants are nodes and the controllers are edges. The network of nonlinear NI plants and the network of OSNI controllers are proved to be a nonlinear NI system and a nonlinear OSNI system, respectively. Under reasonable assumptions, output feedback consensus is established for the networked heterogeneous nonlinear NI systems, and the result is robust against variations in the system models of both the plants and controllers provided that the relevant nonlinear NI and nonlinear OSNI properties are preserved. Finally, an example is given to demonstrate the proposed result on a consensus problem to which earlier results are not applicable.

\bibliographystyle{IEEEtran}
\bibliography{reference.bib}

\begin{thebibliography}{10}
\providecommand{\url}[1]{#1}
\csname url@samestyle\endcsname
\providecommand{\newblock}{\relax}
\providecommand{\bibinfo}[2]{#2}
\providecommand{\BIBentrySTDinterwordspacing}{\spaceskip=0pt\relax}
\providecommand{\BIBentryALTinterwordstretchfactor}{4}
\providecommand{\BIBentryALTinterwordspacing}{\spaceskip=\fontdimen2\font plus
\BIBentryALTinterwordstretchfactor\fontdimen3\font minus
  \fontdimen4\font\relax}
\providecommand{\BIBforeignlanguage}[2]{{%
\expandafter\ifx\csname l@#1\endcsname\relax
\typeout{** WARNING: IEEEtran.bst: No hyphenation pattern has been}%
\typeout{** loaded for the language `#1'. Using the pattern for}%
\typeout{** the default language instead.}%
\else
\language=\csname l@#1\endcsname
\fi
#2}}
\providecommand{\BIBdecl}{\relax}
\BIBdecl

\bibitem{lanzon2008stability}
A.~Lanzon and I.~R. Petersen, ``Stability robustness of a feedback
  interconnection of systems with negative imaginary frequency response,''
  \emph{IEEE Transactions on Automatic Control}, vol.~53, no.~4, pp.
  1042--1046, 2008.

\bibitem{petersen2010feedback}
I.~R. Petersen and A.~Lanzon, ``Feedback control of negative-imaginary
  systems,'' \emph{IEEE Control Systems Magazine}, vol.~30, no.~5, pp. 54--72,
  2010.

\bibitem{bhikkaji2011negative}
B.~Bhikkaji, S.~R. Moheimani, and I.~R. Petersen, ``A negative imaginary
  approach to modeling and control of a collocated structure,'' \emph{IEEE/ASME
  Transactions on Mechatronics}, vol.~17, no.~4, pp. 717--727, 2011.

\bibitem{xiong2009lossless}
J.~Xiong, I.~R. Petersen, and A.~Lanzon, ``On lossless negative imaginary
  systems,'' in \emph{2009 7th Asian Control Conference}.\hskip 1em plus 0.5em
  minus 0.4em\relax IEEE, 2009, pp. 824--829.

\bibitem{song2012negative}
Z.~Song, A.~Lanzon, S.~Patra, and I.~R. Petersen, ``A negative-imaginary lemma
  without minimality assumptions and robust state-feedback synthesis for
  uncertain negative-imaginary systems,'' \emph{Systems \& Control Letters},
  vol.~61, no.~12, pp. 1269--1276, 2012.

\bibitem{xiong2012finite}
J.~Xiong, I.~R. Petersen, and A.~Lanzon, ``Finite frequency negative imaginary
  systems,'' \emph{IEEE Transactions on Automatic Control}, vol.~57, no.~11,
  pp. 2917--2922, 2012.

\bibitem{song2012robust}
Z.~Song, A.~Lanzon, S.~Patra, and I.~R. Petersen, ``Robust performance analysis
  for uncertain negative-imaginary systems,'' \emph{International Journal of
  Robust and Nonlinear Control}, vol.~22, no.~3, pp. 262--281, 2012.

\bibitem{wang2015robust}
J.~Wang, A.~Lanzon, and I.~R. Petersen, ``Robust cooperative control of
  multiple heterogeneous negative-imaginary systems,'' \emph{Automatica},
  vol.~61, pp. 64--72, 2015.

\bibitem{xiong2010negative}
J.~Xiong, I.~R. Petersen, and A.~Lanzon, ``A negative imaginary lemma and the
  stability of interconnections of linear negative imaginary systems,''
  \emph{IEEE Transactions on Automatic Control}, vol.~55, no.~10, pp.
  2342--2347, 2010.

\bibitem{brogliato2007dissipative}
B.~Brogliato, R.~Lozano, B.~Maschke, and O.~Egeland, ``Dissipative systems
  analysis and control,'' \emph{Theory and Applications}, vol.~2, 2007.

\bibitem{bhikkaji2009fast}
B.~Bhikkaji and S.~Moheimani, ``Fast scanning using piezoelectric tube
  nanopositioners: A negative imaginary approach,'' in \emph{2009 IEEE/ASME
  International Conference on Advanced Intelligent Mechatronics}.\hskip 1em
  plus 0.5em minus 0.4em\relax IEEE, 2009, pp. 274--279.

\bibitem{fanson1990positive}
J.~Fanson and T.~K. Caughey, ``Positive position feedback control for large
  space structures,'' \emph{AIAA journal}, vol.~28, no.~4, pp. 717--724, 1990.

\bibitem{cai2010stability}
C.~Cai and G.~Hagen, ``Stability analysis for a string of coupled stable
  subsystems with negative imaginary frequency response,'' \emph{IEEE
  Transactions on Automatic Control}, vol.~55, no.~8, pp. 1958--1963, 2010.

\bibitem{van2010modal}
J.~van Hulzen, G.~Schitter, P.~Van~den Hof, and J.~Van~Eijk, ``Modal actuation
  for high bandwidth nano-positioning,'' in \emph{Proceedings of the 2010
  American Control Conference}.\hskip 1em plus 0.5em minus 0.4em\relax IEEE,
  2010, pp. 6525--6530.

\bibitem{devasia2007survey}
S.~Devasia, E.~Eleftheriou, and S.~R. Moheimani, ``A survey of control issues
  in nanopositioning,'' \emph{IEEE Transactions on Control Systems Technology},
  vol.~15, no.~5, pp. 802--823, 2007.

\bibitem{sebastian2005design}
A.~Sebastian and S.~M. Salapaka, ``Design methodologies for robust
  nano-positioning,'' \emph{IEEE Transactions on Control Systems Technology},
  vol.~13, no.~6, pp. 868--876, 2005.

\bibitem{dong2007robust}
J.~Dong, S.~M. Salapaka, and P.~M. Ferreira, ``Robust {MIMO} control of a
  parallel kinematics nano-positioner for high resolution high bandwidth
  tracking and repetitive tasks,'' in \emph{2007 46th IEEE Conference on
  Decision and Control}.\hskip 1em plus 0.5em minus 0.4em\relax IEEE, 2007, pp.
  4495--4500.

\bibitem{mabrok2014generalizing}
M.~A. Mabrok, A.~G. Kallapur, I.~R. Petersen, and A.~Lanzon, ``Generalizing
  negative imaginary systems theory to include free body dynamics: Control of
  highly resonant structures with free body motion,'' \emph{IEEE Transactions
  on Automatic Control}, vol.~59, no.~10, pp. 2692--2707, 2014.

\bibitem{ghallab2018extending}
A.~G. Ghallab, M.~A. Mabrok, and I.~R. Petersen, ``Extending negative imaginary
  systems theory to nonlinear systems,'' in \emph{2018 IEEE Conference on
  Decision and Control (CDC)}.\hskip 1em plus 0.5em minus 0.4em\relax IEEE,
  2018, pp. 2348--2353.

\bibitem{shi2020robust}
K.~Shi, I.~Vladimirov, and I.~Petersen, ``Robust output feedback consensus for
  networked identical nonlinear negative-imaginary systems,'' \emph{To appear
  in the 24th International Symposium on Mathematical Theory of Networks and
  Systems, arXiv:2005.11492}, 2020.

\bibitem{murray2007recent}
R.~M. Murray, ``Recent research in cooperative control of multivehicle
  systems,'' 2007.

\bibitem{ren2005survey}
W.~Ren, R.~W. Beard, and E.~M. Atkins, ``A survey of consensus problems in
  multi-agent coordination,'' in \emph{Proceedings of the 2005, American
  Control Conference, 2005.}\hskip 1em plus 0.5em minus 0.4em\relax IEEE, 2005,
  pp. 1859--1864.

\bibitem{ren2007consensus}
W.~Ren, ``Consensus strategies for cooperative control of vehicle formations,''
  \emph{IET Control Theory \& Applications}, vol.~1, no.~2, pp. 505--512, 2007.

\bibitem{jadbabaie2003coordination}
A.~Jadbabaie, J.~Lin, and A.~S. Morse, ``Coordination of groups of mobile
  autonomous agents using nearest neighbor rules,'' \emph{IEEE Transactions on
  automatic control}, vol.~48, no.~6, pp. 988--1001, 2003.

\bibitem{olfati2004consensus}
R.~Olfati-Saber and R.~M. Murray, ``Consensus problems in networks of agents
  with switching topology and time-delays,'' \emph{IEEE Transactions on
  automatic control}, vol.~49, no.~9, pp. 1520--1533, 2004.

\bibitem{moreau2005stability}
L.~Moreau, ``Stability of multiagent systems with time-dependent communication
  links,'' \emph{IEEE Transactions on automatic control}, vol.~50, no.~2, pp.
  169--182, 2005.

\bibitem{tanner2003stable}
H.~G. Tanner, A.~Jadbabaie, and G.~J. Pappas, ``Stable flocking of mobile
  agents, part i: Fixed topology,'' in \emph{42nd IEEE International Conference
  on Decision and Control (IEEE Cat. No. 03CH37475)}, vol.~2.\hskip 1em plus
  0.5em minus 0.4em\relax IEEE, 2003, pp. 2010--2015.

\bibitem{saber2003flocking}
R.~O. Saber and R.~M. Murray, ``Flocking with obstacle avoidance: Cooperation
  with limited communication in mobile networks,'' in \emph{42nd IEEE
  International Conference on Decision and Control (IEEE Cat. No. 03CH37475)},
  vol.~2.\hskip 1em plus 0.5em minus 0.4em\relax IEEE, 2003, pp. 2022--2028.

\bibitem{yu2013distributed}
W.~Yu, W.~Ren, W.~X. Zheng, G.~Chen, and J.~L{\"u}, ``Distributed control gains
  design for consensus in multi-agent systems with second-order nonlinear
  dynamics,'' \emph{Automatica}, vol.~49, no.~7, pp. 2107--2115, 2013.

\bibitem{su2011adaptive}
H.~Su, G.~Chen, X.~Wang, and Z.~Lin, ``Adaptive second-order consensus of
  networked mobile agents with nonlinear dynamics,'' \emph{Automatica},
  vol.~47, no.~2, pp. 368--375, 2011.

\bibitem{petersen2016negative}
I.~R. Petersen, ``Negative imaginary systems theory and applications,''
  \emph{Annual Reviews in Control}, vol.~42, pp. 309--318, 2016.

\bibitem{tran2017formation}
V.~P. Tran, M.~Garratt, and I.~R. Petersen, ``Formation control of multi-uavs
  using negative-imaginary systems theory,'' in \emph{2017 11th Asian Control
  Conference (ASCC)}.\hskip 1em plus 0.5em minus 0.4em\relax IEEE, 2017, pp.
  2031--2036.

\bibitem{qi2016cooperative}
Y.~Qi, J.~Wang, Q.~Jia, and J.~Shan, ``Cooperative assembling using multiple
  robotic manipulators,'' in \emph{2016 35th Chinese Control Conference
  (CCC)}.\hskip 1em plus 0.5em minus 0.4em\relax IEEE, 2016, pp. 7973--7978.

\bibitem{hu2019distributed}
J.~Hu, P.~Bhowmick, and A.~Lanzon, ``Distributed adaptive time-varying group
  formation tracking for multiagent systems with multiple leaders on directed
  graphs,'' \emph{IEEE Transactions on Control of Network Systems}, vol.~7,
  no.~1, pp. 140--150, 2019.

\bibitem{peng2016robust}
J.~Peng, J.~Wang, and J.~Shan, ``Robust cooperative tracking of multiple
  p-order power integrators,'' in \emph{2016 Chinese Control and Decision
  Conference (CCDC)}.\hskip 1em plus 0.5em minus 0.4em\relax IEEE, 2016, pp.
  951--956.

\bibitem{skeik2019cooperative}
O.~Skeik, J.~Hu, F.~Arvin, and A.~Lanzon, ``Cooperative control of integrator
  negative imaginary systems with application to rendezvous multiple mobile
  robots,'' in \emph{2019 12th International Workshop on Robot Motion and
  Control (RoMoCo)}.\hskip 1em plus 0.5em minus 0.4em\relax IEEE, 2019, pp.
  15--20.

\bibitem{shi2020robustb}
K.~{Shi}, I.~G. {Vladimirov}, and I.~R. {Petersen}, ``Robust output feedback
  consensus for networked heterogeneous nonlinear negative-imaginary systems,''
  in \emph{2020 Australian and New Zealand Control Conference (ANZCC)}, 2020,
  pp. 214--219.

\bibitem{bhowmick2017lti}
P.~Bhowmick and S.~Patra, ``On {LTI} output strictly negative-imaginary
  systems,'' \emph{Systems \& Control Letters}, vol. 100, pp. 32--42, 2017.

\bibitem{bhowmick2019output}
P.~Bhowmick and A.~Lanzon, ``Output strictly negative imaginary systems and its
  connections to dissipativity theory,'' in \emph{2019 IEEE 58th Conference on
  Decision and Control (CDC)}.\hskip 1em plus 0.5em minus 0.4em\relax IEEE,
  2019, pp. 6754--6759.

\end{thebibliography}

\end{document}